\newcommand*{\wn}{cm$^{-1}$}
\newcommand{\etal}{\emph{et al.}}
\newcommand{\abin}{\emph{ab initio}}
\newcommand{\xstate}{$X^{1}\Sigma_{g}^{+}$}
\newcommand{\efstate}{$EF^{1}\Sigma^{+}_{g}$}
\newcommand*{\Hm}{H$_{2}$}
\newcommand*{\Dm}{D$_{2}$}
\newcommand*{\EFX}{$EF\,{}^{1}\Sigma_{g}^{+}-X\,{}^{1}\Sigma_{g}^{+}$}
\newcommand{\circmark}[1][blue]{\protect\tikz{\protect\draw[#1] (0,0) circle (0.5ex);}}
\newcommand{\rectmark}[1][red]{\protect\tikz{\protect\draw[#1] (0,0) rectangle (0.15,0.15);}}
\newcommand{\trimark}[1][black]{\protect\tikz{\protect\draw[#1] (0,0) -- (0.075,0.15) -- (0.15,0) -- cycle;}}
\newcommand{\circflmark}{\protect\tikz{\protect\draw[black,fill=black] (0,0) circle (0.5ex);}}
\newcommand{\rectflmark}{\protect\tikz{\protect\draw[black,fill=black] (0,0) rectangle (0.15,0.15);}}
\newcommand{\diaflmark}{\protect\tikz{\protect\draw[black,fill=black] (0,0.075) -- (0.075,0.15) -- (0.15,0.075) -- (0.075,0) -- cycle;}}
\def\ao{Appl.\  Opt.\ }

\def\anp{Ann.\ Phys.\ }

\def\apj{Astrophys.\ J. }

\def\jcp{J. Chem.\ Phys.\ }

\def\josa{J. Opt.\ Soc.\ Am.\ }

\def\mp{Mol.\ Phys.\ }

\def\ol{Opt.\ Lett.\ }

\def\pr{Phys.\ Rev.\ }
\def\pra{Phys.\ Rev.\ A }

\def\prd{Phys.\ Rev.\ D }

\def\prl{Phys.\ Rev.\ Lett.\ }

\def\plb{Phys.\ Lett.\ B }

\def\rmp{Rev.\ Mod.\ Phys.\ }
\def\rsi{Rev.\ Sci.\ Instr.\ }
\def\cpl{Chem.\ Phys.\ Lett.\ }

\def\pccp{Phys.\ Chem.\ Chem.\ Phys.\ }
\def\jctc{J.\ Chem.\ Theory\ Comput.\ }
\def\jms{J. Mol.\ Spectrosc.\ }

\def\jpc{J. Phys.\ Chem.\ }

\def\cjp{Can.\ J. Phys.\ }
\def\cjc{Can.\ J. Chem.\ }
\def\zp{Z.\ Phys.\ }
\documentclass[preprint,3p]{elsarticle}
\usepackage{dcolumn}
\usepackage[version=3]{mhchem}
\usepackage[english]{babel}
\usepackage{dcolumn}
\usepackage{graphicx}
\usepackage{amsmath}
\usepackage{gensymb}
\usepackage{tikz}
\usepackage{tabularx}
\usepackage{bigstrut}
\usepackage{longtable}
\usepackage{rotating}
\usepackage{threeparttable}
\usepackage{braket}
\usepackage{verbatim}

\begin{document}

\title{Precision spectroscopy of the $X\,^{1}\Sigma_{g}^{+}, v=0\rightarrow 1$ ($J=0-2$)\\rovibrational splittings in  H$_{2}$, HD and D$_{2}$}

\author[vua]{M.~L.~Niu}
\author[vua,usc]{E.~J.~Salumbides}
\author[vua]{G.~D.~Dickenson\fnref{fn1}}
\author[vua]{K.~S.~E.~Eikema}
\author[vua]{W.~Ubachs\corref{cor}}

\address[vua]{Department of Physics and Astronomy, LaserLaB, VU University, de Boelelaan 1081,\\1081HV Amsterdam, The Netherlands}
\address[usc]{Department of Physics, University of San Carlos, Cebu City 6000, Philippines}
\cortext[cor]{Corresponding author: w.m.g.ubachs@vu.nl}
\fntext[fn1]{Present address: ARC Centre of Excellence for Coherent X-ray Science, School of Chemistry, The University of Melbourne, Victoria 3010, Australia.}

\date{\today}

\begin{abstract}

Accurate experimental values for the vibrational ground tone or fundamental vibrational energy splitting of H$_2$, HD, and D$_2$ are presented. Absolute accuracies of $2\times10^{-4}$ cm$^{-1}$ are obtained from Doppler-free laser spectroscopy applied in a collisionless environment.
The vibrational splitting frequencies are derived from the combination difference between separate electronic excitations from the $X^{1}\Sigma_{g}^{+}, v=0, J$ and $v=1, J$ vibrational states to a common $EF^{1}\Sigma^{+}_{g}, v=0, J$ state.
The present work on rotational quantum states $J=1,2$ extends the results reported by Dickenson~\emph{et al.} on $J=0$ [Phys. Rev. Lett. 110 (2013) 193601].
The experimental procedures leading to this high accuracy are discussed in detail.
A comparison is made with full \emph{ab initio} calculations encompassing Born-Oppenheimer energies, adiabatic and non-adiabatic corrections,
as well as relativistic corrections and QED-contributions.
The present agreement between the experimental results and the calculations provides a stringent test on the application of quantum electrodynamics in molecules.
Furthermore, the combined experimental-theoretical uncertainty can be interpreted to provide bounds to new interactions beyond the Standard Model of Physics or \emph{fifth forces} between hadrons.

\end{abstract}

\begin{keyword}
molecular hydrogen \sep fundamental vibration \sep UV spectroscopy \sep test of QED
\end{keyword}

\maketitle

\section{Introduction}

The first modern explanation of the chemical bond between two neutral hydrogen atoms was put forth in 1927 by Heitler and London~\cite{Heitler1927}, and is one of the earliest applications of quantum theory, specifically that of Schr\"{o}dinger's wave mechanics formulation in 1926~\cite{Schrodinger1926}. Heitler and London showed that by accounting for Pauli's exclusion principle~\cite{Pauli1925} in combining atomic hydrogen wavefunctions to construct molecular wavefunctions, the existence of a bound molecular state is explained. Despite their calculated binding energy being off by some 30\% from the contemporary experimental value, their pioneering quantum mechanical calculation for the stability of molecular hydrogen ushered the era of quantum chemistry. It is interesting to note that the Born-Oppenheimer approximation~\cite{Born1927} was also proposed in 1927, and this approach of separating electronic and nuclear motions has largely shaped molecular theory since.
The next breakthrough in \abin\ potential calculations for H$_2$ was achieved by James and Coolidge in 1933 in their treatment of the ($X^1\Sigma^+_\mathrm{g}$) ground state \cite{James1933}. Using two-electron wave functions with explicitly correlated electrons, an approach introduced by Hylleraas for the helium atom~\cite{Hylleraas1929}, they transcended the concept of electrons being in individual states as used in the Hartree-Fock method. The James-Coolidge solution relied on the variational method to determine the correct nonlinear parameters in combining the wave functions. With a set of only 13 of these wave functions, taken as a truncated basis to represent the total Hilbert space of infinite dimension, they improved the minimum energy in the Born-Oppenheimer potential of the $X\,^1\Sigma^+_\mathrm{g}$ state to $38\,300$ \wn. This was a substantial improvement of about $5\,500$ \wn\ with respect to the best theoretical values available at the time.
Over the years improvements on the accuracy has been obtained~\cite{James1936,Kolos1964,Kolos1965,Kolos1966,Kolos1968,Wolniewicz1993}, and important methodical reviews can be found in Refs.~\cite{Kolos1960,Kolos1963,Kolos1970,Bishop1980,Rychlewski1999}.
The achievement of the initial studies of James and Coolidge~\cite{James1933} can best be appreciated considering that further improvement in the calculated potential has been only 222~\wn\ since then, obtained by Wolniewicz in 1995 with essentially the same method but with a basis of 883 wave functions \cite{Wolniewicz1995}. At present, the Born-Oppenheimer potential energy can be evaluated to accuracies better than 15 digits using more than 22,000 basis functions \cite{Pachucki2010a}, made possible by developments in numerical procedures and improvements in computing power.

\begin{figure}
\centering
\includegraphics[width=0.5\columnwidth]{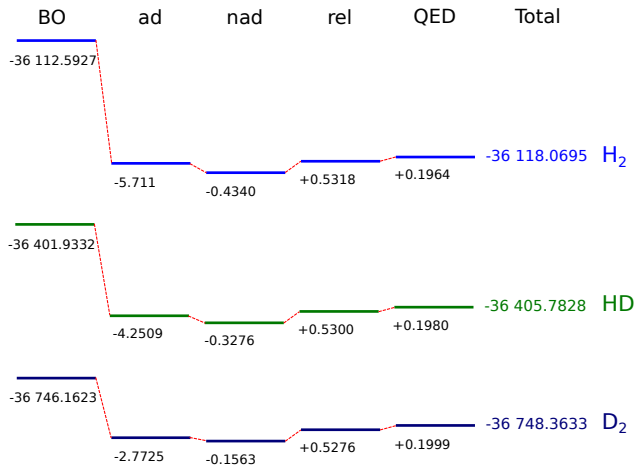}
\caption{(Color online) Graphical representation of the level energy contributions (in \wn) as corrections to the Born-Oppenheimer approximation level energy, with respect to the dissociation limit, of the \xstate, $v=0, J=0$ state for \Hm, HD and \Dm. BO: Born-Oppenheimer energy; ad: adiabatic; nad: nonadiabatic; rel: relativistic; QED: radiative corrections.}
\label{EnergyContribution}
\end{figure}

The precision of the calculated Born-Oppenheimer energy may be considered exact for the purpose of comparisons with experiment. Corrections beyond the Born-Oppenheimer approximation need to be evaluated to improve upon the accuracy of the \abin\ values. In addition to adiabatic and nonadiabatic effects comprising the non-relativistic Born-Oppenheimer corrections, it is also necessary to account for accurate relativistic and radiative or quantum electrodynamic (QED) corrections. Until up to 2010, the work of Wolniewicz~\cite{Wolniewicz1995} that included estimates of radiative corrections, had constituted the state-of-the-art for calculations of level energies in the $X^1\Sigma^+_\mathrm{g}$ ground state of molecular hydrogen. This led to a calculated energy of the actual ground state (or equivalently the dissociation limit) to an accuracy 0.01~\wn. The recent work of Pachucki, Komasa and co-workers has achieved breakthroughs in the evaluation of nonadiabatic effects~\cite{Pachucki2008,Pachucki2009} as well as relativistic and radiative corrections~\cite{Pachucki2005, Pachucki2007}, resulting in accurate level energies of $X^1\Sigma^+_\mathrm{g}$ rovibrational levels~\cite{Piszczatowski2009,Pachucki2010b,Komasa2011}. In Fig.~\ref{EnergyContribution} the different contributions to the level energy of the lowest quantum state ($X^1\Sigma^+_\mathrm{g}, v=0, J=0$) with respect to the dissociation energy of molecular hydrogen are represented graphically to give an impression of the scale of the corrections.

Theoretical and experimental efforts on the determination of ground state level energies in molecular hydrogen mutually stimulated improvements on both fronts as soon as more accurate values were obtained. As an illustration, consider the dissociation energy of the $X^1\Sigma^+_\mathrm{g}$ ground state, a benchmark quantity for the comparison of experiment and theory. The experimental determination of the dissociation energy by Witmer~\cite{Witmer1926} in 1926 already gave results within 3$\%$ of the modern value, an order of magnitude better than the Heitler-London calculations as mentioned. The James-Coolidge calculations in 1933~\cite{James1933} resulted in a dissociation energy that is within $10^{-4}$ of the present value, matched later by the experimental determination by Beutler in 1935~\cite{Beutler1935} that was also accurate to within $10^{-4}$. This lively dynamics continued through the 1960s-1970s, between the experimental efforts of Herzberg and co-workers~\cite{Herzberg1969,Herzberg1970,Herzberg1972} and theoretical efforts by Ko\l{}os and co-workers~\cite{Kolos1965,Kolos1968}.
In the middle of the 1990s, the theoretical result of Wolniewicz~\cite{Wolniewicz1995} for the dissociation energy was expressed in 8 significant digits, although the uncertainty was not explicitly mentioned.
Eyler and co-workers determined the dissociation limit to an accuracy of 0.01 \wn~\cite{Zhang2004} in 2004, improving upon their previous result~\cite{Eyler1993} using the same method.
The most accurate experimental dissociation energy for \Hm\ was obtained in 2009 by Liu~\etal~\cite{Liu2009}, and later extended to \Dm~\cite{Liu2010} and HD~\cite{Sprecher2010}. Remarkably, accurate theoretical values for \Hm\ and \Dm\ dissociation energies~\cite{Piszczatowski2009} as well as HD~\cite{Pachucki2010b} were presented a short time thereafter.

\emph{Ab initio} theory can also be tested through a comparison with the experimental determinations of level splittings in the rovibrational manifold of the ground state. Herzberg first predicted, in 1938, that it should be possible to record rovibrational transitions in the ground state manifold~\cite{Herzberg1938}, and later discovered the quadrupole spectrum in 1949 by photographing a total of eight lines in the (2,0) and (3,0) bands~\cite{Herzberg1949}.
Subsequently the quadrupole spectrum including the fundamental (1,0) band was investigated by several other groups, for example by Rank and co-workers~\cite{Fink1965,Foltz1966}.
The measurements by Bragg~\etal~\cite{Bragg1982} greatly improved the accuracy of the spectroscopy of the quadrupole bands and was until recent years considered as the most accurate work on the direct measurement of the vibrational splittings.
Laser-based direct excitation of the weaker (4,0) and (5,0) overtone quadrupole bands was performed in the visible domain~\cite{Ferguson1993}.
Later investigations using cavity-ring down spectroscopy on the \Hm\ (3,0) overtone band were carried out by Robie~\etal~\cite{Robie2006} using a pulsed source and Hu~\etal~\cite{Hu2012a, Hu2012b} using a cw source.
Campargue and co-workers have recently performed high-resolution determinations of the (2,0) overtone bands of \Hm~\cite{Kassi2012} and \Dm~\cite{Campargue2012} using quantum cascade lasers.
Maddaloni~\etal~\cite{Maddaloni2010} performed precision measurements using cavity-ring down techniques for the fundamental band of \Dm.

The quadrupole excitations in the ground electronic state described above have very low transition probabilities.
The ground state energy splittings can be determined indirectly from appropriate combinations of dipole-allowed transitions between ground state and excited electronic states.
For example, the strongest molecular hydrogen transitions in Lyman ($B^{1}\Sigma_{u}^{+}$ --$X$) and Werner ($C^{1}\Pi_{u}$--$X$) bands have been used to derive ground state rovibrational constants.
Using this approach, Stanke~\etal~\cite{Stanke2008} derived accurate ground state molecular constants based largely on the experimental data of Dabrowski~\cite{Dabrowski1984} but also including quadrupolar transitions.
The natural linewidths of transitions in the Lyman and Werner bands ultimately limit the accuracy that can be achieved~\cite{Philip2004,Ubachs2004}.

In contrast, the rovibrational levels of the lowest-lying excited singlet gerade state \efstate\ of molecular hydrogen have longer natural lifetimes, even up to 150 ns \cite{Chandler1986}, since one-photon transitions to the ground state are forbidden. The gerade states can be accessed from the ground state through two-photon spectroscopies, which also allow for more accurate level energy determinations.
This first excited singlet gerade state in molecular hydrogen, the $EF\,^1\Sigma^+_\mathrm{g}$ state, shown to correspond to a double-well potential~\cite{Davidsson1960},  has been investigated thoroughly over the years.
Eyler and coworkers performed a number of laser spectroscopic studies of increasing accuracy~\cite{Zhang2004,Eyler1987,Eyler1992,Shiner1993,Yiannopoulou2006}.
A determination of frequencies of Q-branch transitions in the lowest \EFX\ (0,0) band was performed with improved accuracy by Hannemann~\etal~\cite{Hannemann2006}.
The lowest rotational levels in the $EF\,^1\Sigma^+_\mathrm{g}$ state derived from the latter study were used as anchor lines, to which a large number of levels in the excited state manifold, obtained from high-resolution Fourier-transform studies, were connected to the ground state \cite{Salumbides2008,Bailly2010}.
Accurate values for level energies of the high rotational states up to $J=16$ in the $E\,^1\Sigma^+_\mathrm{g}, v=0$ electronic state were obtained in Ref.~\cite{Salumbides2011} using UV two-photon spectroscopy.

In this paper, we present accurate experimental and theoretical values for the fundamental vibrational splitting of \Hm, \Dm\ and HD. This extends a recent report~\cite{Dickenson2013} on the \emph{rotationless} ground tone frequencies of hydrogen and its isotopomers, now also including values for $J=1$ and $2$ levels. The experimental determination of the fundamental vibrational splitting is based on combination differences of the transition frequencies between the \xstate\ and \efstate\ states, measured by two-photon Doppler-free spectroscopy.

\begin{figure}
\centering
\includegraphics[width=0.5\columnwidth]{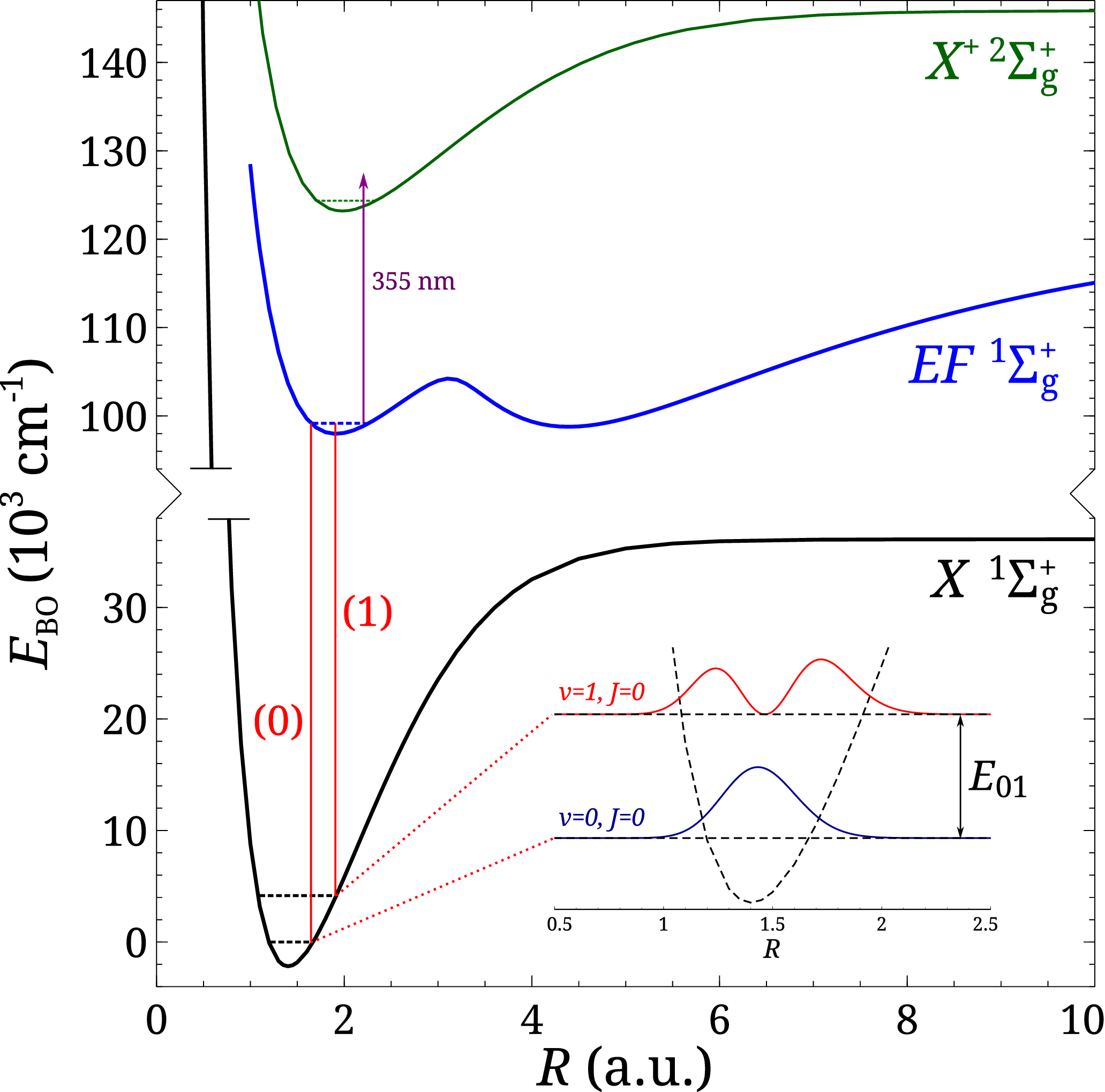}
\caption{(Color online) Potential energy diagram of molecular hydrogen, showing the relevant electronic states accessed in the present spectroscopic investigation. The indirect determination of the fundamental vibrational splitting $E_{01}$ relies on the measurement of Q-line transitions the \EFX\ (0,1) band obtained in the present study and on the measurement of the Q-lines in the \EFX\ (0,0) band, obtained in Ref.~\cite{Hannemann2006}. The excitation channels are indicated by (1) and (0), respectively. An auxiliary laser beam of 355-nm radiation was used in the REMPI detection scheme. The squared moduli of the vibrational wavefunctions are indicated in the inset.}
\label{PotentialEnergy}
\end{figure}

\section{Experiment}

\begin{figure}
\centering
\includegraphics[width=0.6\columnwidth]{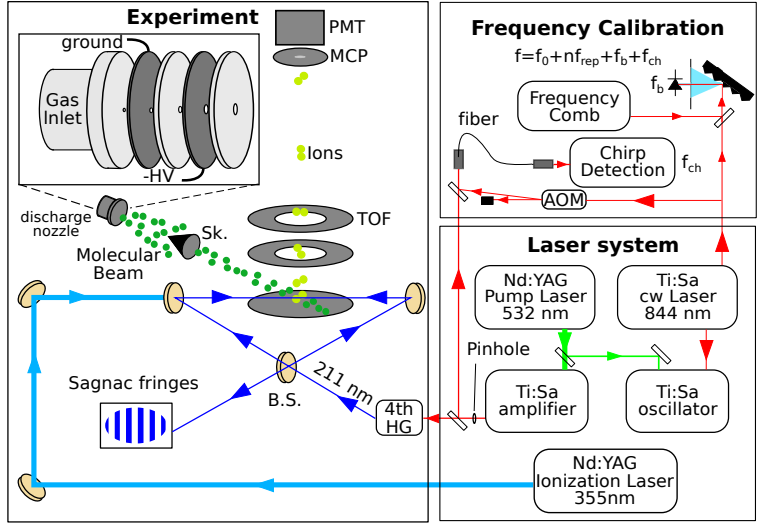}
\caption{(Color online) A schematic layout of the experimental setup, with the laser system, the frequency calibration setup and the molecular beam apparatus with the counter-propagating laser beams in a Sagnac configuration for Doppler-free two-photon spectroscopy. B.S.: beam splitter; Sk.: skimmer. See text for further details.}
\label{setup}
\end{figure}

In this study, high-precision UV two-photon spectroscopy is performed on vibrationally excited molecular hydrogen to determine the transition frequencies of the Q(0), Q(1) and Q(2) lines in the $EF\,{}^{1}\Sigma_{g}^{+} - X\,{}^{1}\Sigma_{g}^{+}\, (0,1)$ band for all three isotopomers H$_2$, HD and D$_2$.
In combination with the previous determination by Hannemann \etal~\cite{Hannemann2006} for the three Q-branch lines in the $EF\,{}^{1}\Sigma_{g}^{+}- X\,{}^{1}\Sigma_{g}^{+}\, (0,0)$ band, accurate values of the fundamental ground tone splittings are obtained for $J=0, 1$ and $2$ rotational levels.
The excitation scheme of the present and previous measurements is drawn in Fig.~\ref{PotentialEnergy}, with the potential energy curves of the relevant electronic states depicted, and the probed two-photon transitions indicated.
The experimental setup is schematically shown in Fig.~\ref{setup}, with blocks representing the narrowband laser source, the frequency calibration setup and the molecular beam machine in which the Doppler-free spectroscopy is performed.

\subsection{Narrowband laser source}

A schematic representation of the laser system is shown as part of Figure~\ref{setup}. The Ti:Sa pulsed laser system is based on an injection-seeded oscillator-amplifier scheme. A continuous wave (cw) Ti:Sa laser serves as the injection-seed for the pulse generation, and is also used to lock the length of the oscillator cavity by a H\"{a}nsch-Couillaud scheme.
The oscillator cavity is pumped with $\sim8$ mJ of 5-ns pulses of the 532-nm (second-harmonic) output from an injection-seeded Nd:YAG laser operating at 10-Hz repetition rate. The resulting pulse from the oscillator cavity is further amplified in a 9-pass bowtie Ti:Sa amplifier pumped with $\sim200$ mJ of the same Nd:YAG pump laser. Typical output pulse energies of the amplifier are approximately 45 mJ at the fundamental wavelength of 844 nm. The pulse duration of the fundamental IR pulses is $\sim$20 ns corresponding to a Fourier-limited bandwidth of $\sim22$ MHz.
An extensive description of the laser system and its operation can be found in Ref.~\cite{Hannemann2007b}, while its performance in the measurements on the \EFX\ (0,0) band were described in Ref.~\cite{Hannemann2006}.
A 1-mm diameter pinhole is used as a spatial filter for the output of the Ti:Sa oscillator-amplifier system before the subsequent harmonic conversion.
Spatial filtering selects a smaller portion of the beam thereby reducing frequency chirp effects across the beam profile, which is crucial in subsequent frequency calibrations.
The spatially-filtered pulsed output is frequency up-converted in two successive frequency-doubling stages (using $\beta$-Barium Borate (BBO) crystals) to yield fourth-harmonic UV radiation, with approximately 350 $\mu$J pulse energies at 211 nm.

\subsection{Two-photon Doppler-free REMPI}

Two-photon Doppler-free techniques were combined with resonantly enhanced multi-photon ionization (REMPI) in the spectroscopic experiment.
In an isotropic gas sample, two-photon absorption from two counter-propagating laser beams results in the cancellation of first-order Doppler shifts.
However, the application in a molecular beam with a defined unidirectional trajectory, results in residual first-order Doppler shifts if there is a misalignment between the counter-propagating laser beams. To improve the laser beam alignment, an interferometric scheme is implemented~\cite{Hannemann2007a}.
The UV probe beam is split in two arms of equal intensity and arranged as part of a Sagnac interferometer, as depicted in Fig.~\ref{setup}, with the interference fringes indicating the degree of alignment.
Narrow bandwidth UV radiation at 211 nm was used to probe the two-photon transitions in \Hm, while HD and \Dm\ measurements required 209 and 207 nm, respectively. A 355-nm laser pulse is used to further ionize the molecules excited in the \efstate, with the ionization laser delayed by 30 ns with respect to the probe laser.
The H$_{2}^{+}$ molecular ions are accelerated by electrostatic lenses and then traverse a field-free time-of-flight (TOF) region. The ions impinge upon a multi-channel plate (MCP) detector attached to a phosphor screen, with the resulting fluorescence collected onto a photomultiplier tube (PMT), for signal registration.

\subsection{Discharge excitation}

The preparation of vibrationally excited molecules in \xstate\ $v=1, J$ is achieved through electron bombardment of the molecular beam in a discharge source similar to that employed in Ref.~\cite{Zhao2011}. The discharge source is a ceramic pinhole nozzle with metallic electrodes (depicted in the top left corner of Fig. \ref{setup}), attached to a pulsed solenoid valve (General Valve Series 9) operated at 10-Hz. The molecular beam pulse is discharged by applying a high voltage pulse to the cathode, with the discharge electrons moving upstream against the molecular beam trajectory. The anode is kept at the same potential as the valve orifice to avoid disturbing the valve operation.
Sufficient population of vibrationally excited molecules was achieved at a voltage of approximately -750 V applied to the cathode.
To reduce ions produced in the same discharge from reaching the detection zone, a pair of deflection plates was installed near the end of the discharge nozzle.
The vibrationally excited molecular hydrogen passes through a skimmer (2-mm diameter) before entering the interaction zone where the UV spectroscopy takes place.

\subsection{Frequency calibration}

The fundamental frequency $f_\mathrm{IR}$ is calibrated by referencing part of the light from the Ti:Sa cw-seed laser to a (Menlo Systems M-comb femtosecond fiber) frequency comb acting as an optical frequency ruler.
The carrier-envelope phase offset frequency $f_{0}$ and repetition frequency $f_{rep}$ of the frequency comb are locked to a local Rubidium-clock that is referenced to the global positioning system.
A heterodyne beat note $f_{b}$ is made between the cw-Ti:Sa laser and the frequency comb modes on an avalanche photodiode, which is counted electronically.
In practice, a number of frequency comb modes participate in the heterodyne process contributing to background noise. Thus, the frequency comb spectrum is dispersed with a grating and subsequent spatial filtering of the unwanted modes is implemented to increase the signal-to-noise ratio in the $f_{b}$ measurement.
The optical frequency of the UV laser system $f_\mathrm{UV}$ can be expressed as
\begin{equation*}
f_\mathrm{UV} = 4 \times f_\mathrm{IR} = 4 \times ( n f_{rep} + f_{0} + f_{b} + f_{ch} ),
\end{equation*}
where the prefactor of 4 accounts for the harmonic order, and $n$ is mode number of the frequency comb component used. The cw-pulse frequency offset $f_{ch}$ of the Ti:Sa system due to frequency chirp will be discussed below. The mode number determination follows from a coarse calibration of the laser using a Burleigh wavemeter accurate to $\sim$30 MHz, which is sufficient for an unambiguous mode assignment since $f_{rep}\sim250$ MHz. The sign of the respective frequency contributions to $f_\mathrm{IR}$ may be positive or negative but can be easily determined in practice.

\subsection{Ti:Sa cw-pulse frequency offset}

In the frequency calibration procedure, the Ti:Sa cw-seed optical frequency $f_\mathrm{IR}$ is determined, while the output of the oscillator-amplifier Ti:Sa \textit{pulsed} laser system is actually used in the spectroscopy, after frequency upconversion to the 4$^{th}$ harmonic.
Any frequency offset $f_{ch}$ between the Ti:Sa cw-seed frequency and pulsed output frequency needs to be determined, where it is possible that the optical frequency is time-dependent within the laser pulse.
The pulse generated in the Ti:Sa oscillator is subject to cavity-mode pulling effects,
as a result of an optically-induced change in refractive index of the Ti:Sa crystal due to the intense 532-nm pump pulse, contributing to a frequency offset.
The actual cavity resonance in the presence of the pump pulse is then frequency-shifted with respect to the locking point of the unpumped cavity.
The frequency shift can be separated into contributions from a thermal change of the refractive index $\Delta n_\mathrm{th}$ and an effect from the population inversion $\Delta n_\mathrm{inv}$ in the Ti:Sa crystal~\cite{Hannemann2007b}.
The total effective frequency offset, typically tens of MHz in the fundamental $f_\mathrm{IR}$, can be compensated by controlling the lock setpoint of the Ti:Sa oscillator, e.g. locking at the side of the fringe in the H\"{a}nsch-Couillaud scheme.

A related phenomenon is spatial frequency chirp where the pulsed optical frequency varies across the transverse beam profile of the laser beam. This effect occurs in the amplification stage, because subsequent passes sample a different area within the pump beam profile that has a Gaussian intensity distribution. The spatial frequency difference from either edges of the beam profile was found to be a few MHz with respect to $f_\mathrm{IR}$. To minimize such a spatial frequency offset, a 1-mm pinhole after the Ti:Sa amplifier acts as a spatial filter while still providing sufficient energy for the frequency upconversion process.

The cw-pulse offset frequency $f_{ch}$ is measured by heterodyning the amplified pulsed output of the Ti:Sa system with part of the cw-seed that is shifted by 250 MHz using an acousto-optic modulator (AOM).
The resulting beat signal from a fast photo-detector is recorded with an oscilloscope for further frequency chirp analysis~\cite{Fee1992,Eikema1997}.
A single-shot frequency chirp analysis is performed for every Ti:Sa pulse in order to account for the frequency offset $f_{ch}$ during each measurement point.
A more detailed description of the frequency chirp measurements and analysis for the Ti:Sa pulsed laser system is given in Refs.~\cite{Hannemann2006,Hannemann2007b}.

\begin{figure}
\centering
\includegraphics[width=0.5\columnwidth]{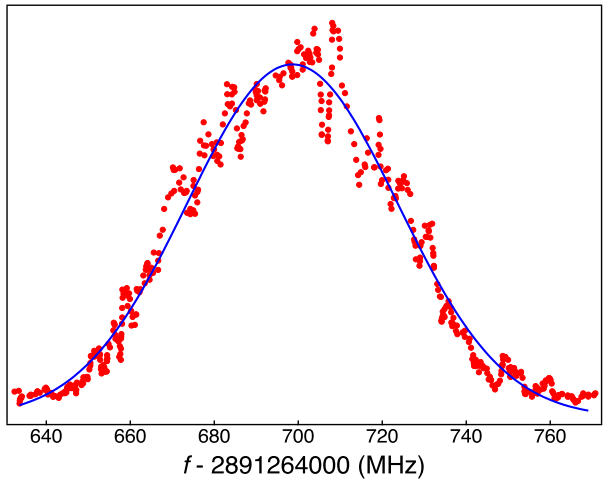}
\caption{(Color online) Recording of the $EF^{1}\Sigma_{g}^{+}-X^{1}\Sigma_{g}^{+}$(0,1) Q(1) two-photon transition in D$_{2}$. A Gaussian profile (blue line) fitted to the averaged datapoints (in red) is shown.}
\label{spect}
\end{figure}

\subsection{Assessment of systematic effects}

A spectral recording of the \Dm\ $EF^{1}\Sigma_{g}^{+}-X^{1}\Sigma_{g}^{+}$(0,1) Q(1) two-photon transition is shown in Fig.~\ref{spect}.
The full-width half-maximum of the spectral lines is $\sim60$ MHz and is well-approximated by a Gaussian profile, limited by the instrumental linewidth of the UV laser source, and accounting for the two-photon excitation.
As a test of robustness, the obtained line positions were also checked when imposing different background levels.
The uncertainty contribution is then estimated to be better than 500 kHz for most scans of good signal-to-noise ratio.
For more noisy spectra, the line fitting uncertainty contribution is difficult to separate from statistical scatter, thus it is subsumed into the statistics entry.
Several measurements of a specific transition taken over several days, demonstrate reproducibility of within $\sim2$ MHz in the transition frequency.

\subsubsection{ac-Stark effect}

An important systematic effect is the ac-Stark frequency shift induced by the power density of the probe radiation.
The use of a separate ionization laser, delayed with respect to the probe, allows for sufficient detection efficiency even at a reduced probe beam intensity, thereby minimizing the ac-Stark shift.
The ac-Stark measurements are performed in two different beam configurations, referred to as (I) and (II). For most of the measurements, the UV beam is collimated with a beam diameter of 0.5 mm, using a combination of concave mirrors, that also serves to reduce beam astigmatism.
This configuration (I) is typical for low UV power density.
An additional benefit of configuration (I) is that it reduces residual first-order Doppler shifts compared to the case when using counter-propagating beams with curved wavefronts~\cite{Yiannopoulou2006}.
Using (I), a 355-nm laser pulse, delayed by 30 ns relative to the probe beam, was employed in order to ionize molecular hydrogen in the \efstate\ state.
To aid in the ac-Stark shift assessment and to improve the extrapolation to the field-free case, a second configuration is implemented, referred to as (II). In this configuration, a lens with 1-m focal length is placed before the beam splitter in Fig.~\ref{setup}, thus reducing the beam diameter to $\sim80$~$\mu$m in the interaction region so that the power density is increased by a factor $\sim 40$ with respect to that of (I).
The intensity of the spectroscopy UV beam in (II) is sufficient to induce ionization, so that a separate ionization laser is not required.

\begin{figure}
\centering
\includegraphics[width=0.5\columnwidth]{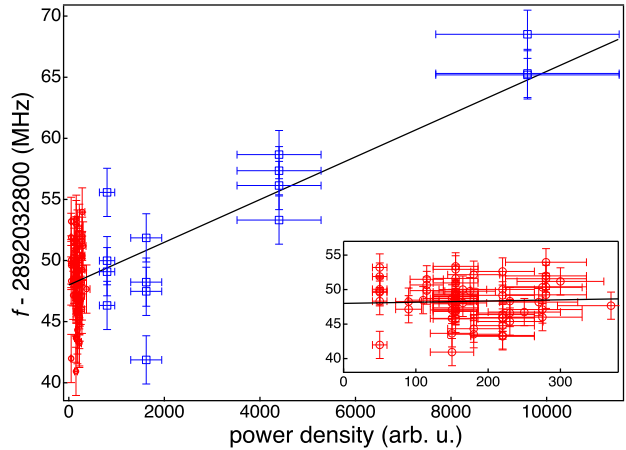}
\caption{(Color online) Assessment of the ac-Stark effect for the \EFX\ (0,1) Q(0) transition in D$_{2}$ plotted against the 207-nm probe power density in the interaction zone. Blue datapoints were collected using focused probe beams (Conf. II), while red datapoints were taken with collimated probe beams (Conf. I) as also shown in the inset.}
\label{stark}
\end{figure}

Measurements are performed for different probe beam intensities in both configurations and the results are illustrated in Fig.~\ref{stark}.
The measurements for higher power densities (II) are shown in the full graph, while the measurements at low power density (I) are enlarged in the inset.
The field-free transition frequency at zero intensity is obtained from a weighted linear fit of the combined measurement results of (I) and (II).
The horizontal axis represents the relative power density and is derived from a measurement of the probe beam intensity and a measurement of the beam diameter at the interaction region.
It is worth noting that for the energy range between 30 and 300 $\mu$J in (I), the ac-Stark frequency shifts are within statistical scatter.
The estimated error from the ac-Stark effect is deduced from the error of the intercept determined in the weighted linear fit, with the extrapolated field-free frequency estimated to be accurate to $\sim 0.4$ MHz.
This extrapolation procedure was performed for every transition measured.

\subsubsection{dc-Stark effect}
The dc-Stark effect is avoided by pulsing the voltages of the ion extraction plates so that the transitions are probed under dc field-free conditions. However, no measurable change in the transition frequencies was observed when the extraction fields were operated in either pulsed- or dc-mode, and we therefore estimate a contribution of $\leq 0.1$ MHz on the systematic uncertainty due to the dc Stark effect.

\subsubsection{Laser beam alignment}
\begin{figure}
\centering
\includegraphics[width=0.5\columnwidth]{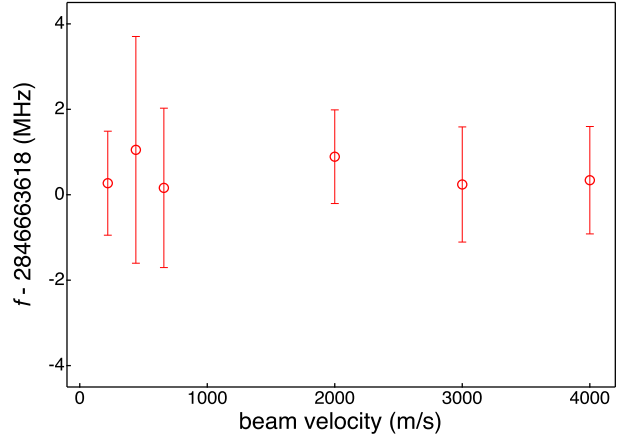}
\caption{(Color online) Transition frequency of the \Hm\ Q(1) line measured at different molecular beam rms speed, obtained from different concentrations of \Hm-Kr mixtures, and time delay settings between valve opening and laser trigger.}
\label{Doppler}
\end{figure}

The residual first-order Doppler shifts estimated from the Sagnac interferometer alignment of the counter-propagating probe beams were experimentally verified by purposely misaligning the probe beams, where the resulting shifts were found to be below statistical scatter.
Further tests were also performed by using mixed samples of molecular hydrogen and krypton to reduce the speed of \Hm.
More finely-tuned molecular beam velocities were obtained by varying the delay between the timing of the valve opening and the trigger of the laser pulse, by means of which molecules in the leading, middle and trailing parts of the gas pulse are probed. 
Fig.~\ref{Doppler} shows the frequency measurements of the \EFX\ (0,1) Q(1) line for the six different \Hm\ beam velocities thus obtained.
No effect above the statistical uncertainty is observed.

\subsubsection{Pressure shifts}
To assess possible pressure shifts, we use the pressure shift coefficient for Rydberg states from Ref.~\cite{Herzberg1972} of $5.7 (5)$ \wn/amagat as an upper limit, noting that for the pure ground electronic state, quadrupole transition shifts are in the order of $\sim3\times 10^{-3}$ \wn/amagat, see e.g.~\cite{Kassi2012}.
From local gas densities used in the experiment, an upper limit for the pressure shift of $0.06$ MHz is estimated.

\subsection{Uncertainty estimates}

Table~\ref{Tab:Uncertainty} summarizes the uncertainty contributions in MHz from different error sources in the determination of the \EFX\ (0,1) transitions.
Systematic corrections were applied separately for each spectral recording, for example for the frequency chirp and ac-Stark shift.
The statistics entries denote the statistical $1\sigma$ standard deviations of all the measurements after the contributions of various systematic effects have been corrected for.
The uncertainties in the determination of the line positions are not indicated as they are already included in the statistical entry, where the averaging is weighted by uncertainty in the line fitting for each recorded spectrum.
The data collection for each line covered a period of different days, where the different transitions were remeasured throughout the whole measurement period to confirm the reproducibility of the results.

\begin{table}
\caption{Estimated systematic and statistical uncertainty contributions for the frequency calibrations of the \EFX\ (0,1) transitions in \Hm , HD and \Dm. The uncertainty values (Unc.) are given in MHz. }
\label{Tab:Uncertainty}
\begin{threeparttable}
\begin{tabular}{ll@{\hspace{15pt}}c@{\hspace{15pt}}r}
\\
\hline\hline\noalign{\vskip 2mm}
\multicolumn{2}{c}{Contribution}&Species&Unc.\\[1mm]
\hline\noalign{\vskip 2mm}
(i) 	&ac-Stark\tnote{a}	& 	&$<0.4$\\
(ii) 	&dc-Stark		&	&$<0.1$\\
(iii) 	&frequency chirp\tnote{a}&	&2.0\\
(iv) 	&frequency calibration	&	&0.1 \\
(v) 	&residual 1st-order Doppler	&\Hm	&0.5\\
 	&			&HD	&0.3\\
 	&			&\Dm	&0.3\\
(vi) 	&2nd-order Doppler	&	&$<0.1$\\
(vii)	&pressure shift		&	&$<0.1$\\

(viii) 	&statistics		&\Hm	&1.5\tnote{b}\\
 	&			&HD	&1.6\tnote{c}\\
 	&			&\Dm	&1.9\tnote{d}\\[2mm]
\multicolumn{2}{l}{Total Uncertainty\tnote{e}}	&\Hm 	&2.6\\
	&					&HD	&2.6\\
	&					&\Dm	&2.8\\[1mm]
\hline\hline
\end{tabular}
\begin{tablenotes}
\footnotesize
\item[a] ac-Stark and chirp offsets are corrected for and not indicated in the table.
\item[b] Standard deviation based on 63 measurements.
\item[c] Standard deviation based on 69 measurements.
\item[d] Standard deviation based on 64 measurements.
\item[e] Quadrature sum of errors.
\end{tablenotes}
\end{threeparttable}
\end{table}

\section{Results}

Q-branch transition frequencies in the \EFX\ (0,1) band are listed in Table~\ref{EFXtransitions} for \Hm, HD, \Dm. The Q(1) and Q(2) transitions for each isotopomer extend the rotationless transitions reported by Dickenson~\etal~\cite{Dickenson2013}. Also listed in Table~\ref{EFXtransitions} are the corresponding transition frequencies in the \EFX\ (0,0) band taken from Hannemann~\etal~\cite{Hannemann2006}. The improved uncertainties for the present \EFX\ (0,1) transitions are primarily due to the reduction in the spatial frequency chirp of the Ti:Sa laser system output by spatial filtering. The data on the \EFX\ (0,1) band presented here are in agreement with results of Eyler~\etal~\cite{Eyler1987}, with the present results representing a 100-fold improvement over the previous measurement.

\begin{table*}
\caption{Transition frequencies for Q-lines in the  \EFX\ (0,1) band in \Hm , HD and \Dm.
Data for the \EFX\ (0,0) band from Ref.~\cite{Hannemann2006}. All values in \wn.}
\begin{center}
\label{EFXtransitions}
\begin{tabular}{c@{\hspace{15pt}}c@{\hspace{15pt}}c@{\hspace{15pt}}c}
\hline\hline\noalign{\vskip 2mm}	
	&		Q(0)		&	Q(1)		&	Q(2)		\\	
\noalign{\vskip 2mm}\hline
\noalign{\vskip 2mm}	
\multicolumn{4}{c}{\EFX(0,1)}	\\
\noalign{\vskip 2mm}
\Hm\	&	95\,003.620\,55\,(10)	&	94\,954.477\,39\,(10)	&	94\,856.717\,48\,(10)	\\			
HD	&	95\,669.186\,10\,(10)	&	95\,631.613\,43\,(10)	&				\\
\Dm\	&	96\,467.832\,02\,(10)	&	96\,442.209\,32\,(10)	&	96\,391.100\,00\,(10)	\\
\noalign{\vskip 3mm}
\multicolumn{4}{c}{ \EFX(0,0)}	\\
\noalign{\vskip 2mm}
\Hm\	&	99\,164.786\,91\,(11)		&	99\,109.731\,39\,(18)	&	99\,000.183\,01\,(11)\\			
HD	&	99\,301.346\,62\,(20)		&	99\,259.917\,93\,(20)	&	\\
\Dm\	&	99\,461.449\,08\,(11)		&	99\,433.716\,38\,(11)	&	99\,378.393\,52\,(11)\\
\noalign{\vskip 2mm}\hline\hline
\end{tabular}
\end{center}
\end{table*}

\emph{Ab initio} values for the vibrational energy splittings in the \xstate\ electronic ground state were presented in Refs.~\cite{Piszczatowski2009,Komasa2011} for \Hm\ and \Dm\, while the rovibrational level energies of the HD ground state were given in Ref.~\cite{Pachucki2010b}.
These state-of-art calculations are based on Nonadiabatic Perturbation Theory (NAPT) \cite{Pachucki2008,Pachucki2009} to obtain the nonrelativistic energy contributions, while the nonrelativistic quantum electrodynamics (NRQED) formalism~\cite{Pachucki2005,Caswell1986,Pachucki2006} is used to pertubatively obtain the relativistic and QED energy terms.
More accurate theoretical values for the rotationless transitions of \Hm, HD and \Dm\ were presented in Ref.~\cite{Dickenson2013}. The theoretical vibrational energy splittings are summarized in Table~\ref{Tab:Splittings}, where the various energy contributions are separately listed, including the estimated uncertainties where available. The most accurate values are for the rotationless transitions from Ref.~\cite{Dickenson2013} with accuracies of better than $1\times10^{-4}$ \wn. For the $J=1,2$ transitions of \Hm\ and \Dm, accurate values were obtained from the supplementary material of Komasa \etal~\cite{Komasa2011}, however, only the accuracy of the total level energies was indicated. The obtained fundamental vibrational transition energies are estimated to be accurate to $1\times10^{-3}$ \wn. For the $J=1\rightarrow1$ HD transition, level energies from Pachucki and Komasa~\cite{Pachucki2010b} were used, however, the energy contributions were not separately indicated.
The estimated theoretical uncertainty for the more accurate rotationless transition energies stems from the better uncertainty of the nonadiabatic corrections.
This estimate is aided by available theoretical results for the rotationless vibrational splittings from Adamowic and co-workers~\cite{Stanke2008, Stanke2009, Bubin2010}, using an independent theoretical methodology based on a variational procedure to obtain the nonadiabatic wave functions. Similar calculations for transitions involving $J\neq 0$, however, are not available. Since Komasa, Pachucki and co-workers obtain transition energies using the same method for the rotationless case as for all $J$ quantum numbers, the original uncertainty estimates indicated in Refs.~\cite{Piszczatowski2009,Pachucki2010b,Komasa2011} are likely to be overestimated. 

\begin{table}
\caption{\emph{Ab initio} values for the fundamental vibrational energy splittings ($v=0\rightarrow1$) of the  \xstate\ ground state for the three isotopomers
H$_2$, HD and D$_2$.
The different columns indicate the transitions labeled by the rotational quantum numbers of the states involved.
All values are in \wn\ with 1$\sigma$ uncertainties given in between parentheses ().
For the rotationless transitions, values without indicated uncertainties have negligible contributions, limited by numerical precision.
For the relativistic and QED effects, $\mathcal{R}$ denotes the Rydberg constant and the order in $\alpha$ is indicated for each correction.
The higher-order term HQED also includes estimates of the next order corrections.
For the $J=1\rightarrow1$ and $J=2\rightarrow2$ values the combinations are taken from Refs.~\cite{Piszczatowski2009,Pachucki2010b,Komasa2011} where
binding energies are listed separately for each $(v,J)$ level; here the uncertainties are taken in quadrature from both independent values.
For the $J=0\rightarrow0$ values a cancellation of uncertainties is assumed as discussed in Ref.~\cite{Dickenson2013}.
}
\label{Tab:Splittings}
\begin{center}
\begin{tabular}{l r@{.}l r@{.}l r@{.}l }
\hline
\hline\noalign{\vskip 1mm}
 H$_2$ &\multicolumn{2}{c}{$J=0\rightarrow0$}&\multicolumn{2}{c}{$J=1\rightarrow1$}&\multicolumn{2}{c}{$J=2\rightarrow2$}\\
\noalign{\vskip 1mm}\hline
\noalign{\vskip 2mm}
Born-Oppenheimer	&	4\,163	&	403\,50	&	4\,157	&	483\,7	&	4\,145	&	680\,5	\\
Adiabatic       	&	-1	&	402\,84	&	-1	&	396\,3	&	-1	&	383\,2	\\
Nonadiabatic    	&	-0	&	836\,49	&	-0	&	835\,4	&	-0	&	833\,5	\\
\noalign{\vskip 0.8mm}
Nonrel subtotal ($\alpha^{0}\mathcal{R}$)	&	4\,161	&	164\,16\,(1)	&	4\,155	&	252\,0	&	4\,143	&	463\,8	\\
\noalign{\vskip 2mm}
Relativistic ($\alpha^{2}\mathcal{R}$)  	&	0	&	023\,41\,(3)	&	0	&	023\,2	&	0	&	022\,7	\\
QED ($\alpha^{3}\mathcal{R}$)   	&	-0	&	021\,29\,(2)	&	-0	&	021\,2	&	-0	&	021\,1	\\
HQED ($\alpha^{4}\mathcal{R}$)  	&	-0	&	000\,16\,(8)	&	-0	&	000\,2	&	-0	&	000\,2	\\
\noalign{\vskip 0.8mm}													
Rel + QED subtotal              	&	0	&	001\,96\,(9)	&	0	&	001\,8	&	0	&	001\,4	\\
\noalign{\vskip 2mm}													
Theory total                    	&	4\,161	&	166\,12\,(9)	&	4\,155	&	253\,8\,(9)	&	4\,143	&	465\,3(9)	\\
\noalign{\vskip 1mm}
\hline
\hline
 HD &\multicolumn{2}{c}{$J=0\rightarrow0$}&\multicolumn{2}{c}{$J=1\rightarrow1$}\\
\noalign{\vskip 1mm}\hline
\noalign{\vskip 2mm}
Born-Oppenheimer			&	3\,633	&	719\,56		&\multicolumn{2}{c}{--}\\
Adiabatic       			&	-0	&	932\,59		&\multicolumn{2}{c}{--}\\
Nonadiabatic    			&	-0	&	628\,72		&\multicolumn{2}{c}{--}\\
\noalign{\vskip 0.8mm}
Nonrel subtotal ($\alpha^{0}\mathcal{R}$)&	3\,632	&	158\,26(1)	&\multicolumn{2}{c}{--}\\
\noalign{\vskip 2mm}
Relativistic ($\alpha^{2}\mathcal{R}$)  &	0	&	020\,93(2)	&\multicolumn{2}{c}{--}\\
QED ($\alpha^{3}\mathcal{R}$)   	&	-0	&	018\,63(2)	&\multicolumn{2}{c}{--}\\
HQED ($\alpha^{4}\mathcal{R}$)  	&	-0	&	000\,14(7)	&\multicolumn{2}{c}{--}\\
\noalign{\vskip 0.8mm}								 
Rel + QED subtotal              	&	0	&	002\,16(8)	&\multicolumn{2}{c}{--}\\
\noalign{\vskip 2mm}								   
Theory total                    	&	3\,632	&	160\,41(8)	& 3\,628&304\,4\,(10)\\
\noalign{\vskip 1mm}
\hline
\hline
 D$_2$ &\multicolumn{2}{c}{$J=0\rightarrow0$}&\multicolumn{2}{c}{$J=1\rightarrow1$}&\multicolumn{2}{c}{$J=2\rightarrow2$}\\
\noalign{\vskip 1mm}\hline
\noalign{\vskip 2mm}
Born-Oppenheimer	&	2\,994	&	440\,84	&	2\,992	&	329\,5	&	298\,8	&	113\,3	\\
Adiabatic       	&	-0	&	521\,50	&	-0	&	520\,4	&	-0	&	518\,0	\\
Nonadiabatic    	&	-0	&	304\,47	&	-0	&	304\,3	&	-0	&	304\,0	\\
\noalign{\vskip 0.8mm}
Nonrel subtotal ($\alpha^{0}\mathcal{R}$)	&	2\,993	&	614\,87\,(1)	&	2\,991	&	504\,8	&	2\,987	&	291\,3	\\
\noalign{\vskip 2mm}
Relativistic ($\alpha^{2}\mathcal{R}$)  	&	0	&	017\,71\,(2)	&	0	&	017\,6	&	0	&	017\,5	\\
QED ($\alpha^{3}\mathcal{R}$)   	&	-0	&	015\,39\,(2)	&	-0	&	015\,4	&	-0	&	015\,3	\\
HQED ($\alpha^{4}\mathcal{R}$)  	&	-0	&	000\,12\,(6)	&	-0	&	000\,1	&	-0	&	000\,1	\\
\noalign{\vskip 0.8mm}													
Rel + QED subtotal              	&	0	&	002\,20\,(7)	&	0	&	002\,1	&	0	&	002\,1	\\
\noalign{\vskip 2mm}													
Theory total                    	&	2\,993	&	617\,08\,(7)	&	2\,991	&	507\,0\,(2)	&	2\,987	&	293\,4\,(2)	\\
\noalign{\vskip 1mm}
\hline
\hline
\end{tabular}
\end{center}
\end{table}

Ground state \xstate\ energy splittings are obtained from the combination differences of the \EFX\ transition energies and are listed in Table~\ref{Vibtransitions}.
The indicated uncertainty of the fundamental vibrational energy splittings is the quadrature sum of the uncertainties in the particular \EFX\ transition energies used.
The difference between the experimental and theoretical results $\Delta E$ is also listed in Table~\ref{Vibtransitions}, along with the combined uncertainty $\delta E = \sqrt{\delta E_\mathrm{exp}^{2} + \delta E_\mathrm{the}^{2} }$, from the experimental $\delta E_\mathrm{exp}$ and theoretical $\delta E_\mathrm{the}$ uncertainties.
The accuracy of present experimental values for the vibrational transitions is predominantly limited by the \EFX\ (0,0) results from Hannemann \etal~\cite{Hannemann2006}. For the rotationless transitions the theoretical values are more accurate than the experimental ones. However, for transitions involving $J=1,2$ the experimental results are 5 times more accurate.
For the transitions listed, the difference $\Delta E$ is statistically consistent with null within the combined uncertainty $\delta E$.
(The $1.5\sigma$ deviation for the \Hm\ $J=0\rightarrow 1$ comparison is compatible with expectations from statistics.)
The overall comparison demonstrates excellent agreement between the present experimental and theoretical values.

\begin{table*}
\caption{Fundamental vibrational energy splittings $(v=0\rightarrow 1)$ in \Hm , HD and \Dm. The third column is the difference between the experimental and theoretical values, $\Delta E = E_\mathrm{exp} - E_\mathrm{the}$, while $\delta E$ represents the combined experimental and theoretical uncertainty. All values in \wn.}
\begin{center}
\label{Vibtransitions}
\begin{tabular}{c @{\hspace{15pt}} r@{.}l @{\hspace{15pt}} r@{.}l @{\hspace{15pt}} r@{.}l@ {\hspace{15pt}} r@{.}l}

\hline\hline\noalign{\vskip 2mm}	
	&	\multicolumn{2}{c}{Experiment}&\multicolumn{2}{c}{Theory}&\multicolumn{2}{c}{$\Delta E$}&\multicolumn{2}{c}{$\delta E$}\\	
\noalign{\vskip 2mm}\hline

\noalign{\vskip 2mm}	
\multicolumn{9}{c}{$J = 0\rightarrow 0$}	\\
\noalign{\vskip 2mm}

\Hm\	&	4\,161&166\,36\,(15)	&	4\,161&166\,12\,(9)	&	 0&000\,24	&	0&000\,17\\			
HD	&	3\,632&160\,52\,(22)	&	3\,632&160\,41\,(8)	&	 0&000\,11	&	0&000\,23\\
\Dm\	&	2\,993&617\,06\,(15)	&	2\,993&617\,08\,(7)	&	-0&000\,02	&	0&000\,17\\

\noalign{\vskip 3mm}

\multicolumn{9}{c}{$J = 1\rightarrow 1$}	\\
\noalign{\vskip 2mm}

\Hm\	&	4\,155&254\,00\,(21)	&	4\,155&253\,8\,(9)	& 	0&000\,2	&	0&000\,9\\			
HD	&	3\,628&304\,50\,(22)	&	3\,628&304\,4\,(10)	&	0&000\,1	&	0&001\,0\\
\Dm\	&	2\,991&507\,06\,(15)	&	2\,991&507\,0\,(2)	&	0&000\,1	&	0&000\,3\\

\noalign{\vskip 3mm}

\multicolumn{9}{c}{$J = 2\rightarrow 2$}	\\
\noalign{\vskip 2mm}

\Hm\	&	4\,143&465\,53\,(15)	&	4\,143&465\,3\,(9)	&	0&000\,2	&	0&000\,9\\			
\Dm\	&	2\,987&293\,52\,(15)	&	2\,987&293\,4\,(2)	&	0&000\,1	&	0&000\,3\\
\noalign{\vskip 2mm}\hline\hline                                                                         
\end{tabular}
\end{center}
\end{table*}

\section{Comparison to previous studies}

Various methods have been employed in direct excitations of ground state rovibrational transitions that include among others quadrupole absorption studies, Raman spectroscopy and electric-field induced dipole spectroscopy.
Since the first measurements of the quadrupole spectra of \Hm\ overtone bands by Herzberg~\cite{Herzberg1949} using long-path classical absorption techniques, refinements have been applied through the years, e.g. by using more accurate echelle spectrometers by Rank and co-workers~\cite{Fink1965, Foltz1966, Rank1962}. The improvements continued in the study of Bragg~\etal~\cite{Bragg1982} who employed Fourier-Transform (FT) spectroscopy techniques. Since the first measurements of Rasetti~\cite{Rasetti1929} and the early investigations of Stoicheff~\cite{Stoicheff1957}, the Raman spectrum has also been measured with ever increasing accuracies. From the first demonstrations by Crawford and Dagg~\cite{Crawford1953} of the electric field induced dipole excitation, parallel improvements in accuracy have also been achieved by later investigators. The weak transitions probed necessitated high pressures that led to collisional shifts and broadening effects first investigated by May~\etal~\cite{May1961,May1964}.
Numerous other investigations on the \Hm\ fundamental band at varying accuracies for the transition energies include Refs.~\cite{Looi1978, Veirs1987, Germann1988, Rahn1990}.

\begin{figure}
\centering
\includegraphics[width=0.85\columnwidth]{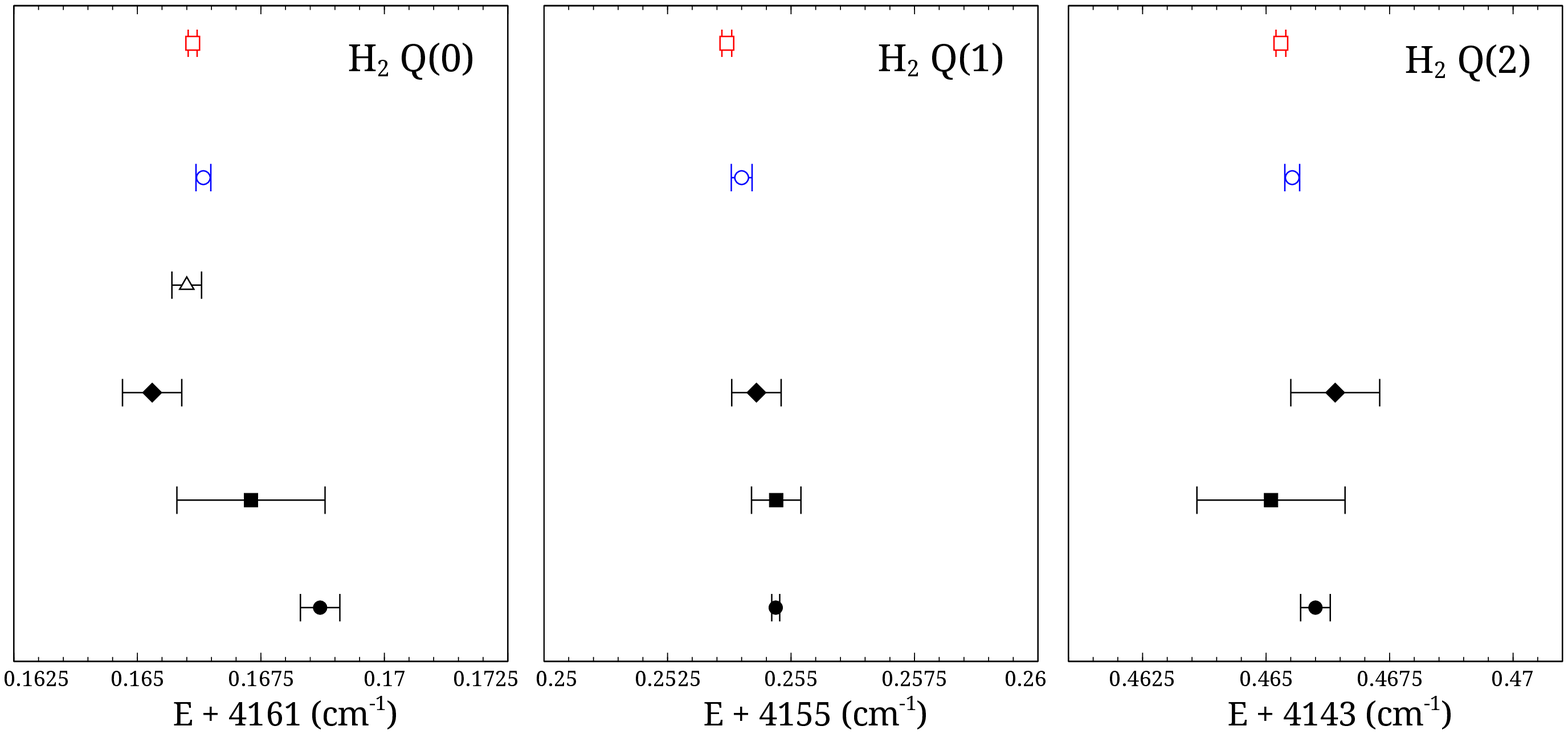}
\caption{
Comparison of the present experimental results (\circmark) for the \Hm\ ground state fundamental band Q-branch transitions to previous studies representing the most accurate values from a particular method:
from long-path absorption reported by Bragg~\etal~\cite{Bragg1982} (\circflmark); from Raman spectroscopy by Rahn and Rosasco~\cite{Rahn1990} (\rectflmark); from spectroscopy of electric field induced transitions by Buijs~\cite{Buijs1971} (\diaflmark).
The results on empirical fitting of the global \Hm\ database by Stanke~\etal~\cite{Stanke2008} (\trimark) are also included, along with the present theoretical results (\rectmark).
}
\label{H2comp}
\end{figure}

The comparison of the present experimental results to selected previous determinations is shown graphically in Fig.~\ref{H2comp} for \Hm.
Only those investigations with the highest claimed accuracy for a particular method are included in the figure.
The results of Bragg and co-workers~\cite{Bragg1982} stood as the most accurate for decades, making use of Fourier Transform spectroscopy with long-path absorption samples.
For the Q(1) transition, however, the result from Bragg \etal\ differs from the present result by several standard deviations. In that study, the collision-induced dipole spectrum adds a broad background signal for the (1,0) band (e.g. Fig.~2 in Ref.~\cite{Bragg1982}) that could affect the determination of the line positions.
The most accurate electric field induced spectrum was recorded by Buijs~\cite{Buijs1971} who also employed Fourier Transform spectroscopy.
For Raman spectroscopy, the most accurate measurements were performed by Rahn and Rosasco~\cite{Rahn1990}, using a pulsed laser source based on difference-frequency mixing.
The results of Rahn and Rosasco~\cite{Rahn1990} and that of Buijs~\cite{Buijs1971} are in fair agreement with the present results.
The latter comparison suggests that electric-field induced frequency shifts are less severe than pressure-induced systematic shifts, even after pressure-shift corrections.

Durie and Herzberg~\cite{Durie1960} recorded the weak dipole absorption spectrum of the (1,0), (2,0), (3,0) and (4,0) bands in the ground state of HD.
Stoicheff~\cite{Stoicheff1957} performed Raman spectroscopy on the fundamental band of HD, while Brannon~\etal~\cite{Brannon1968} measured electric-field induced transitions in the same band.
McKellar and co-workers~\cite{McKellar1973, McKellar1976, Rich1982} carried out several long-path absorption investigations covering the fundamental and overtone bands up to (6,0) band of the HD ground state. In these investigations, the pressure shifts on the transition energies were systemically studied as was the case in Ref.~\cite{Nazemi1983}.
The Raman study of Veirs and Rosenblatt~\cite{Veirs1987} also included the HD fundamental band, 
as well as the fundamental band for most other molecular hydrogen isotopologues.

\begin{figure}
\centering
\includegraphics[width=0.55\columnwidth]{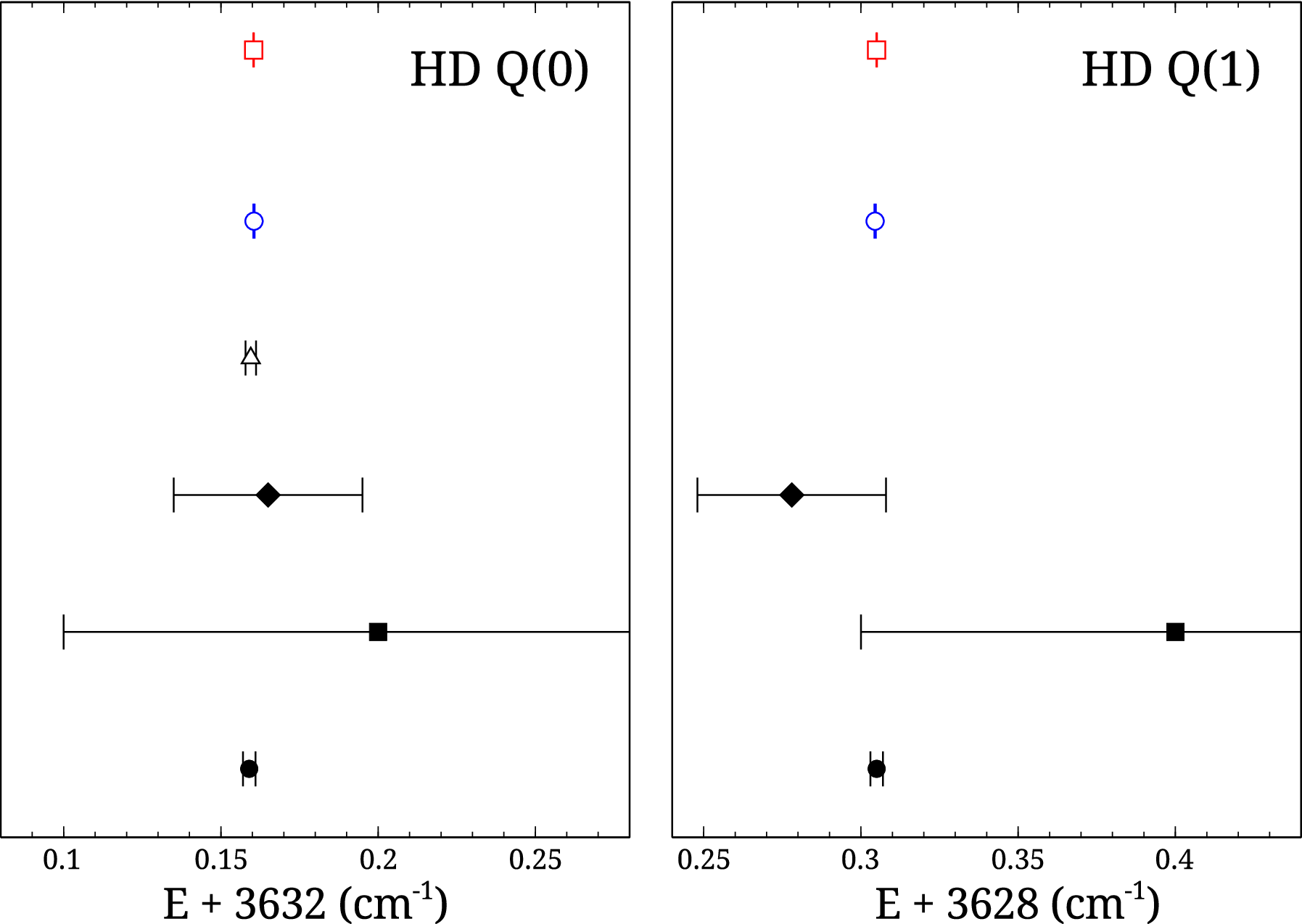}
\caption{
Comparison of the present experimental results (\circmark) for the HD ground state fundamental band Q-branch transitions to previous studies representing the most accurate values from a particular method:
from long-path absorption reported by Rich~\etal~\cite{Rich1982} (\circflmark); from Raman spectroscopy by Veirs and Rosenblatt~\cite{Veirs1987} (\rectflmark); from spectroscopy of electric field induced transitions by Brannon~\etal~\cite{Brannon1968} (\diaflmark).
The results on empirical fitting of the global HD database by Stanke~\etal~\cite{Stanke2009} (\trimark) are also included, along with the present theoretical results (\rectmark).
Note that for this energy scale the error bars for the present experimental and theoretical results are not resolved and thus overlap.
}
\label{HDcomp}
\end{figure}

Fig.~\ref{HDcomp} is a graphical representation of the comparison of the present experimental results for the HD fundamental band to selected previous determinations.
Rich~\etal~\cite{Rich1982} reported the most accurate results with Fourier Transform spectroscopy on long-path absorption samples.
The most accurate electric field induced spectrum was that of Brannon~\etal~\cite{Brannon1968} while Raman spectra for HD were obtained by Veirs and Rosenblatt~\cite{Veirs1987}.

\begin{figure}
\centering
\includegraphics[width=0.85\columnwidth]{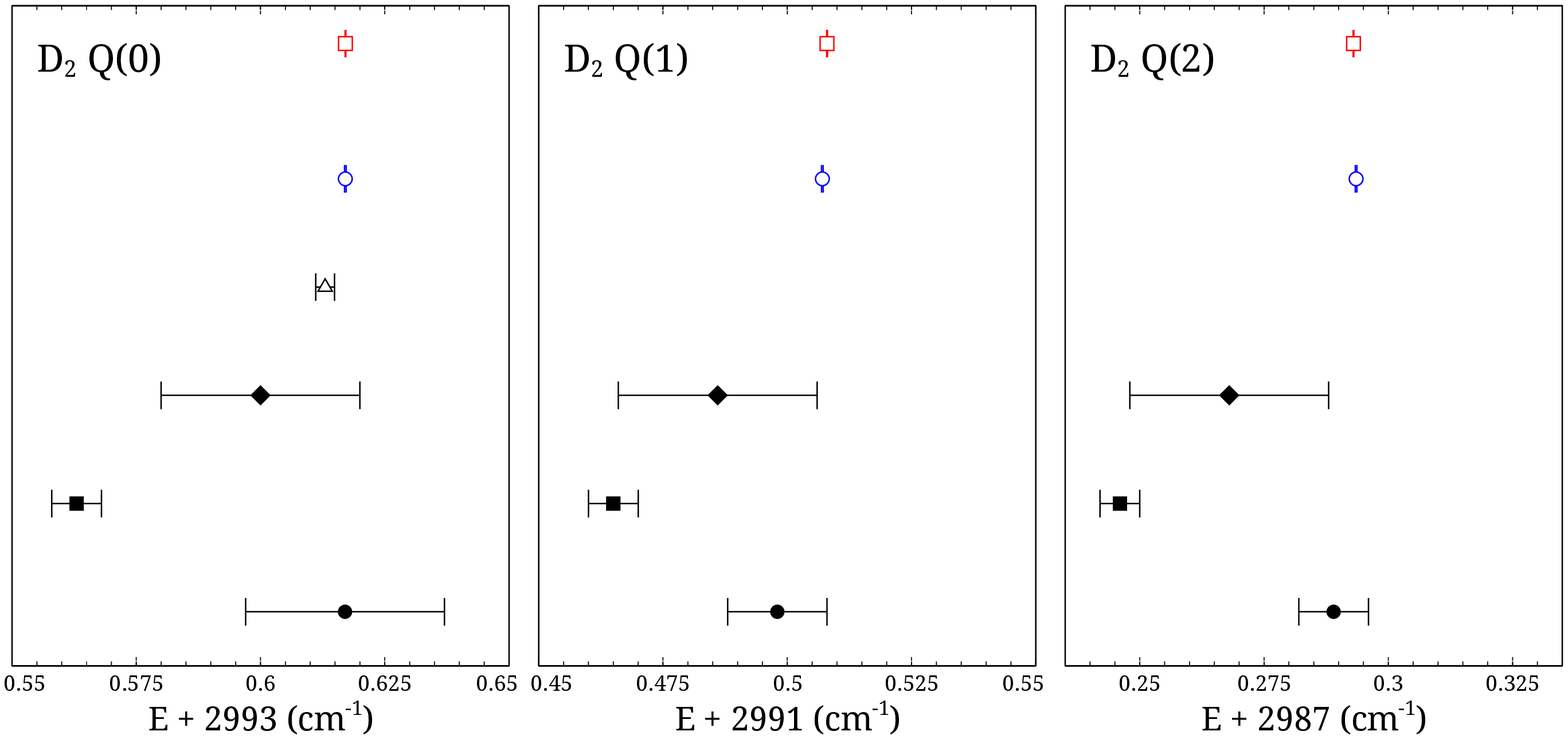}
\caption{
Comparison of the present experimental results (\circmark) for the \Dm\ ground state fundamental band Q-branch transitions to previous studies representing the most accurate values from a particular method:
from long-path absorption reported by Jennings~\etal~\cite{Jennings1986} (\circflmark); from Raman spectroscopy by Looi~\etal~\cite{Looi1978} (\rectflmark); from spectroscopy of electric field induced transitions by Brannon~\etal~\cite{Brannon1968} (\diaflmark).
The results on empirical fitting of the global \Dm\ database by Bubin~\etal~\cite{Bubin2010} (\trimark) are also included, along with the present theoretical results (\rectmark).
Note that for this energy scale the error bars for the present experimental and theoretical results are not resolved and thus overlap.
}
\label{D2comp}
\end{figure}

The fundamental band of \Dm\ was also included in the Raman spectroscopic investigations of Stoicheff~\cite{Stoicheff1957}. Later, Looi~\etal~\cite{Looi1978} improved upon the Raman spectroscopy, and in addition investigated pressure dependent energy shifts. Electric field induced transitions of \Dm\ were also measured by Brannon~\etal~\cite{Brannon1968} with pressure shift corrections. McKellar and Oka~\cite{McKellar1978} used a difference-frequency laser system to investigate the fundamental vibrational band of \Dm\ in a long-path cell. Jennings~\etal~\cite{Jennings1986} performed accurate long-path absorption measurements using FT spectroscopy.
A comparison of the \Dm\ fundamental band transitions is plotted in Fig.~\ref{D2comp}. The older results from long-path absorption and electric-field induced spectra coincide with the present results. The results from the Raman investigations deviate considerably from the other studies including the present, suggesting systematic errors in Ref.~\cite{Looi1978}.

An indirect approach through empirical fitting of Dunham coefficients have been carried out by Stanke~\etal~\cite{Stanke2008} for \Hm. Experimental data included in the fit came from Dabrowski~\cite{Dabrowski1984} for Lyman and Werner band transitions, but also most ground state studies mentioned above~\cite{Foltz1966,Stoicheff1957,Veirs1987,Buijs1971,Brannon1968,Jennings1986}.
Similar global fitting analyses based on the Dunham relation was also performed by the same group for HD~\cite{Stanke2009} and \Dm~\cite{Bubin2010} resulting in similar uncertainties as in the case of \Hm.
Remarkable accuracy was achieved for these global fits, despite considerable deviations of results from individual experiments used.

The excitation of molecular hydrogen in a collisionless environment in the present study is distinct from previous studies, where the low excitation probability required the use of dense gas samples with pressures of a few bars for detection.
The recorded lines had Doppler-widths on the order of GHz and the collisional perturbation gives rise to pressure shifts amounting to 100 MHz~\cite{Campargue2012}. These issues are effectively absent in the present study using molecular beam conditions.
As a case in point, it is interesting to look at the claimed uncertainty of Maddaloni~\etal~\cite{Maddaloni2010}, who deployed a high-resolution laser source based on difference-frequency-generation referenced to the Cs-clock primary standard using an optical frequency comb synthesizer. Nevertheless, a comparison with theory gives differences of $0.0015(3)$ and $-0.0010(3)$ \wn\ for S(0) and S(1) in Ref.~\cite{Maddaloni2010}, respectively, as pointed out in Ref.~\cite{Komasa2011}. The excellent agreement of the present experimental results with the same \abin\ calculations suggests that the systematic uncertainties in Ref.~\cite{Maddaloni2010} might be underestimated.

\section{Testing QED and fifth forces}

As can be seen in Table~\ref{Vibtransitions}, the \abin\ values of the rotationless ground tone energies are the most accurate for all isotopomers. For these energy splittings, the relativistic and QED terms exhibit the dominant uncertainty contribution to the theoretical value while the nonrelativistic energy hardly contributes to the uncertainy. By subtracting the theoretical nonrelativistic energy term from the experimental value, one obtains a hybrid experimental-theoretical determination of the relativistic and QED contribution of $0.002\,20(17)$ \wn\ for \Hm.
Comparing the latter value to the \abin\ values represents a test of relativistic and QED calculations in molecules on the order of $\sim1\%$.
Applying similar arguments for the dissociation limit of the ground electronic state of molecular hydrogen, and using the experimental results of Ref.~\cite{Liu2009} and theoretical values from Ref.~\cite{Piszczatowski2009}, relativistic and QED calculations in molecules are verified on the order of $\sim0.1\%$.

The excellent agreement between theory and experiment discussed above, can be exploited to constrain effects beyond the Standard Model of Physics. Since molecular (and atomic) structure is dominated by the electromagnetic interaction, with the effects of the strong-, weak- and gravitational forces many orders of magnitudes weaker (at least for light systems as hydrogen, where weak force effects are not yet detectable), any discrepancy between measurements and theory points to new physics.

New interactions or modifications beyond the Standard Model are expected as it does not provide explanations for several phenomena, e.g. Dark Matter \cite{Bertone2005} and Dark Energy \cite{Peebles2003}.
The possibility that these hypothetical new interactions or fifth forces have subtle effects in atomic or molecular structure is a complementary approach to the searches and tests in particle physics or astronomy.

Energy resonances in calculable few-body systems provide an ideal search ground. Simple systems to search for extra interactions between lepton and hadrons would be atomic hydrogen or the helium ion~\cite{Karshenboim2010, Ubachs2013}, while extra hadron-hadron interactions can be probed in molecular hydrogen and molecular hydrogen ions.
We express a general fifth force in terms of a Yukawa potential $V_5(R)$ as a function of the distance $R$ between hadrons, or the internuclear distance in molecular hydrogen:
\begin{equation}
	V_5(R) \,=\, \alpha_5 N_1 N_2 \frac{\exp{(-R/\lambda})}{R} \hbar c \,=\, \alpha_5 N_1 N_2 Y_\lambda(R),
\label{Yuk0}
\end{equation}
where $\alpha_5$ is an interaction strength, and $\lambda$ is the characteristic range of the interaction. We assume that the long-range effect (as opposed to the short range of the strong force) scales with the number of nucleons of each nucleus, $N_1$ and $N_2$ respectively, and we treat protons and neutrons equally. The potential $V_5(R)$ can be considered as a perturbation on the energies of quantum states, leading to a differential energy shift $\Braket{ \Delta V_{5,\lambda} }$ on energy level differences between states $(v',J')$ and $(v'', J'')$:
\begin{eqnarray}
	\Braket{ \Delta V_{5,\lambda} }  &=&  \alpha_5 N_{1} N_{2} \left[ \Braket{ \Psi_{v',J'}(R)  |  Y(r,\lambda) | \Psi_{v',J'}(R) } \right. \nonumber \\
                   && \qquad \qquad \left. - \Braket{ \Psi_{v'',J''}(R)  |  Y(r,\lambda) | \Psi_{v'',J''}(R) }  \right] \nonumber \\
					&=& \alpha_5 N_{1} N_{2} \Delta Y_{\lambda}
\label{YV5}
\end{eqnarray}
where $\Psi_{v,J}(R)$ are the wave functions representing the probability of finding the nuclei at a certain separation $R$ within the molecule.
Using accurate wave functions from Ref.~\cite{Piszczatowski2009,Komasa2011}, the quantity $\Delta Y_{\lambda}$ has been evaluated taking $\lambda$ as a parameter in Ref.~\cite{Salumbides2013}.
$\Delta Y_{\lambda}$ can be considered as the inherent sensitivity coefficient for a specific transition, which becomes larger as the wavefunctions of the two levels involved are more different (e.g. wavefunctions in Fig.~\ref{PotentialEnergy}). Thus transitions with greater $\Delta v$ have a larger $\Braket{ \Delta V_{5,\lambda} }$ energy shift. The largest inherent sensitivity is expected for the dissociation energy $D_{0}$ since the shift is
$\Braket{\Delta V_{5,\lambda}}  = - \alpha_5 N_1 N_2 \Braket{\Psi_{v=0}(R)  | Y(r,\lambda) | \Psi_{v=0}(R)}$.
For a particular transition, $\Delta Y_{\lambda}$ is greatest for \Hm\ and least for \Dm\ since the spatial extent of the wavefunctions between any two levels, with corresponding set of quantum numbers, are more similar for \Dm\ than in \Hm. However, the effect of nucleon numbers, $N_1$ and $N_2$, more than compensates for the lower $\Delta Y_{\lambda}$ in the heavier isotopomers, so that the expected energy shift $\Braket{ \Delta V_{5,\lambda} }$ is actually greatest for \Dm\ and least for \Hm\ for a particular transition.

\begin{figure}
\centering
\includegraphics[width=0.65\columnwidth]{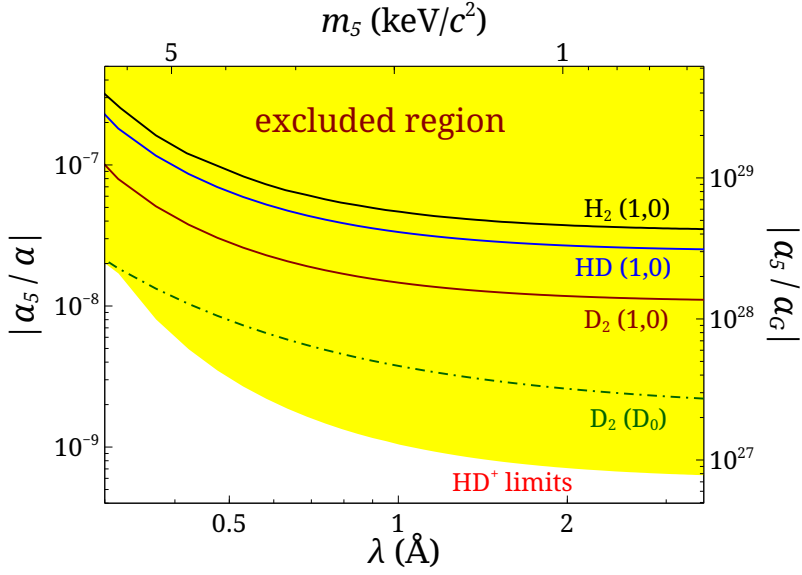}
\caption{
Derived bounds for the interaction strength $\alpha_5$ of possible fifth forces relative to the fine structure constant (left vertical axis) or the gravitational force (right axis), for different interaction length range $\lambda$ obtained from the fundamental vibration of molecular hydrogen. Similar bounds from the  \Dm\ dissociation energy $D_{0}$ as well as rovibrational transitions in HD$^{+}$ are also indicated. The yellow area indicates excluded regions based mainly on constraints obtained from HD$^{+}$.
}
\label{Limits}
\end{figure}

An upper bound for the interaction strength $\alpha_5$ can be be obtained from the combined experimental-theoretical uncertainty $\delta E$ for a particular transition by the relation $\alpha_5 < \delta E / (N_1 N_2 \Delta Y_{\lambda})$. Fig.~\ref{Limits} shows the constraint for the interaction strength $\alpha_5$, for certain range  $\lambda$ of the interaction, obtained from the rotationless fundamental vibrational transitions in \Hm, HD and \Dm, with the shaded area indicating the excluded region. The fifth-force interaction strength is given in terms of the strengths of the electromagnetic force, $\alpha$, and the gravitational force, $\alpha_G$. More stringent bounds in the same length scale are discussed in Ref.~\cite{Salumbides2013} based on spectroscopic determinations on the dissociation energy $D_0$ of molecular hydrogen~\cite{Liu2009, Liu2010, Sprecher2010} as well as rovibrational transitions in HD$^{+}$~\cite{Koelemeij2007, Koelemeij2012, Bressel2012}. Similar constraints have been extended to shorter interaction length at sub-Angstrom scales in Ref.~\cite{Salumbides2014} based on level splittings in exotic molecules. The interaction length scale $\lambda$ can be associated with the mass $m_5$ of a fifth-force carrier particle as indicated in the upper horizontal axis in Fig.~\ref{Limits}. This boson mass range is complementary to the mass sensitivity range in particle accelerators on the high end, and the low-mass sensitivity in gravitational test experiments on the other extreme.

From the present measurements on level splittings of the fundamental ground tones in the hydrogen molecule and its isotopomers, a constraint on the strength of a possible fifth force between hadrons is determined to be $|\alpha_5/\alpha|<2\times 10^{-8}$ for a force range of 1 \AA.

\section{Conclusion}

The fundamental vibrational energy splitting of H$_2$, HD, and D$_2$ were determined to absolute accuracies of $2\times10^{-4}$ cm$^{-1}$, or a relative accuracy of a few parts in $10^{-8}$.
The vibrational splitting frequencies are derived from the combination difference between separate electronic excitations from the $X^{1}\Sigma_{g}^{+}, v=0, J$ and $v=1, J$ vibrational states to a common $EF^{1}\Sigma^{+}_{g}, v=0, J$ state.
Doppler-free laser spectroscopic investigation applied in a collisionless molecular beam environment leads to high accuracy, where pressure effects are negligible in contrast to studies based on gas cells.
The excellent agreement between the experimental results and the calculations provides a stringent test on the application of quantum electrodynamics in molecules, and can be used to provide bounds to new interactions.
Upper bounds derived from molecular hydrogen indicate that the interaction strength of possible fifth forces must be at least 8 orders of magnitude weaker than the electromagnetic strength, for a fifth-force interaction range in the order of typical internuclear distances of $\sim1$ \AA. 
This brings molecular spectroscopy studies again to the forefront of physics, reminiscent of the early days of quantum mechanics.

\smallskip
\section*{Acknowledgments}
This work was supported by the Netherlands Astrochemistry Program of NWO (CW-EW) and by the Netherlands Foundation for Fundamental Research of Matter (FOM) through the program "Broken Mirrors \& Drifting Constants".


\vspace{0.5cm}

\small
\begin{thebibliography}{999}

\bibitem{Heitler1927} W. Heitler and F. London, \zp 44 (1927) 619.
\bibitem{Schrodinger1926} E. Schr\"{o}dinger, \pr 28 (1926) 1049.
\bibitem{Pauli1925} W. Pauli, \zp 31 (1925) 765.
\bibitem{Born1927} M. Born and R. Oppenheimer, \anp 389 (1927) 457.
\bibitem{James1933} H. J. James and A. S. Coolidge, \jcp 1 (1933) 825 .
\bibitem{Hylleraas1929} E. A. Hylleraas, \zp 54 (1929) 347.
\bibitem{James1936} H. J. James and A. S. Coolidge, \pr 49 (1936) 688.
\bibitem{Kolos1964} W. Ko{\l}os and L. Wolniewicz, \jcp 41 (1964) 3663.
\bibitem{Kolos1965} W. Ko{\l}os and L. Wolniewicz, \jcp 43 (1965) 2429. 
\bibitem{Kolos1966} W. Ko{\l}os and L. Wolniewicz, \jcp 45 (1966) 515.
\bibitem{Kolos1968} W. Ko{\l}os and L. Wolniewicz, \jcp 49 (1968) 404.
\bibitem{Wolniewicz1993} L. Wolniewicz, \jcp 99 (1993) 1851.
\bibitem{Kolos1960} W. Ko{\l}os and C. J. J. Roothaan,  Rev.\ Mod.\ Phys.\ 32 (1960) 205.
\bibitem{Kolos1963} W. Ko{\l}os and L. Wolniewicz, Rev.\ Mod.\ Phys.\ 35 (1963) 473.
\bibitem{Kolos1970} W. Ko{\l}os, Adv.\ Quantum Chem.\ 5 (1970) 99.
\bibitem{Bishop1980} D. M. Bishop and L. M. Cheung, Adv.\ Quantum Chem.\ 12 (1980) 1.
\bibitem{Rychlewski1999} J. Rychlewski, Adv.\ Quantum Chem.\ 31 (1999) 173. 
\bibitem{Wolniewicz1995} L. Wolniewicz, \jcp 103 (1995) 1792.
\bibitem{Pachucki2010a} K. Pachucki, \pra 82 (2010) 032509.
\bibitem{Pachucki2008} K. Pachucki and J. Komasa, \jcp 129 (2008) 034102.
\bibitem{Pachucki2009} K. Pachucki and J. Komasa, \jcp 130 (2009) 164113.
\bibitem{Pachucki2005} K. Pachucki, \pra 71 (2005) 012503.
\bibitem{Pachucki2007} K. Pachucki, \pra 76 (2007) 022106.
\bibitem{Piszczatowski2009} K. Piszczatowski, G. \L ach, M. Przybytek, J. Komasa, K. Pachucki and B. Jeziorski, \jctc 5 (2009) 3039.
\bibitem{Pachucki2010b}K. Pachucki and J. Komasa, \pccp 12 (2010) 9188.
\bibitem{Komasa2011} J. Komasa, K. Piszczatowski, G. Lach, M. Przybytek, B. Jeziorski and K. Pachucki, \jctc 7 (2011) 3105.
\bibitem{Witmer1926} E. E. Witmer, \pr\ 28 (1926) 1223.
\bibitem{Beutler1935} H. Beutler, Z. Phys. Chem. Abt. 29 (1935) 315.
\bibitem{Herzberg1969} G. Herzberg, \prl 23 (1969) 1081.
\bibitem{Herzberg1970} G. Herzberg, \jms 33 (1970) 147.
\bibitem{Herzberg1972} G. Herzberg and C. Jungen, \jms 41 (1972) 425.
\bibitem{Zhang2004} Y. P. Zhang, C. H. Cheng, J. T. Kim, J. Stanojevic, and E. E. Eyler, \prl 92 (2004) 203003. 
\bibitem{Eyler1993} E. E. Eyler and N. Melikechi, \pra 48 (1987) R18. 
\bibitem{Liu2009} J. Liu, E. J. Salumbides, U. Hollenstein, J. C. J. Koelemeij, K. S. E. Eikema, W. Ubachs and F. Merkt, \jcp 130 (2009) 174306.
\bibitem{Liu2010} J. Liu, D. Sprecher, C. Jungen, W. Ubachs and F. Merkt, \jcp 132 (2010) 154301.
\bibitem{Sprecher2010}  D. Sprecher, J. Liu, C. Jungen, W. Ubachs and F. Merkt, \jcp 133 (2010) 111102.
\bibitem{Herzberg1938} G. Herzberg, \apj 87 (1938) 428.
\bibitem{Herzberg1949} G. Herzberg, Nature 163 (1949) 170.
\bibitem{Fink1965} U. Fink, T. A. Wiggins and D. H. Rank, \jms 18 (1965) 384.
\bibitem{Foltz1966} J. V. Foltz, D. H. Rank and T. A. Wiggins, \jms 21 (1966) 203 .
\bibitem{Bragg1982} S. L. Bragg, W. H. Smith and J. W. Brault, \apj 263 (1982) 999. 
\bibitem{Ferguson1993} D.W. Ferguson, K.N. Rao, M.E. Mickelson and L.E. Larson, \jms 160 (1993) 315.
\bibitem{Robie2006} D. C. Robie and J. T. Hodges, \jcp 124 (2006) 024307. 
\bibitem{Hu2012a} C.-F. Cheng, Y. R. Sun, H. Pan, J. Wang,  A. W. Liu, A. Campargue and S.-M. Hu, \pra 85 (2012) 024501.
\bibitem{Hu2012b} S.-M. Hu, H. Pan, C.-F. Cheng, Y. R. Sun, X. F. Li, J. Wang, A. Campargue and A. W. Liu, \apj 749 (2012) 76.
\bibitem{Kassi2012} S. Kassi, A. Campargue, K. Pachucki and J. Komasa, \jcp 136 (2012) 184309.
\bibitem{Campargue2012} A. Campargue, S. Kassi, K. Pachucki and J. Komasa, \pccp 14 (2012) 802.
\bibitem{Maddaloni2010} P. Maddaloni, P. Malara, E. De Tommasi, M. De Rosa, I. Ricciardi, G. Gagliardi, F. Tamassia, G. Di Lonardo and P. De Natale, \jcp 133 (2010) 154317.
\bibitem{Stanke2008} M. Stanke, D. Kedziera, S. Bubin, M. Molski and L. Adamowicz, \jms 128 (2008) 114313.
\bibitem{Dabrowski1984} I. Dabrowski, \cjp 62 (1984) 1639.
\bibitem{Philip2004} J. Philip, J. P. Sprengers, Th. Pielage, C. A. de Lange, W. Ubachs and E. Reinhold, \cjc 82 (2004) 713.
\bibitem{Ubachs2004} W. Ubachs and E. Reinhold, \prl 92 (2004) 101302.
\bibitem{Chandler1986} D. W. Chandler and L. R. Thorne, \jcp 85 (1986) 1733.
\bibitem{Davidsson1960} E. R. Davidsson, \jcp 33 (1960) 1577.
\bibitem{Eyler1987} E. E. Eyler, J. Gilligan, E. F. McCormack, A. Nussenzweig and E. Pollack, \pra 36 (1987) 3486.
\bibitem{Eyler1992} J.M. Gilligan and E.E. Eyler, \pra 46 (1992) 3676.
\bibitem{Shiner1993} D. Shiner, J. M. Gilligan, B. M. Cook, and W. Lichten, \pra 47 (1993) 4042.
\bibitem{Yiannopoulou2006} A. Yiannopoulou, N. Melikechi, S. Gangopadhyay, J. C. Meiners, C. H. Cheng and E. E. Eyler, \pra 73 (2006) 022506.
\bibitem{Hannemann2006} S. Hannemann, E. J. Salumbides, S. Witte, R. T. Zinkstok, E.-J. van Duijn, K. S. E. Eikema and W. Ubachs, \pra 74 (2006) 062514.
\bibitem{Salumbides2008} E. J. Salumbides, D. Bailly, A. Khramov, A. L. Wolf, K. S. E. Eikema, M. Vervloet and W. Ubachs, \prl 101 (2008) 223001.
\bibitem{Bailly2010} D. Bailly, E. J. Salumbides, M. Vervloet and W. Ubachs, \mp 108 (2010) 827.
\bibitem{Salumbides2011} E. J. Salumbides, G. D. Dickenson, T. I. Ivanov and W. Ubachs, \prl 107 (2011) 043005.
\bibitem{Dickenson2013} G. D. Dickenson, M. L. Niu, E. J. Salumbides, J. Komasa, K. S. E. Eikema, K. Pachucki and W. Ubachs, \prl 110 (2013) 193601.
\bibitem{Hannemann2007b} S. Hannemann, E.-J. van Duijn and W. Ubachs, \rsi 78 (2007) 103102.
\bibitem{Hannemann2007a} S. Hannemann, E. J. Salumbides and W. Ubachs, \ol 32 (2007) 1381.
\bibitem{Zhao2011} D. Zhao, N. Wehres, H. Linnartz and W. Ubachs, \cpl 501 (2011) 232.
\bibitem{Fee1992} M. S. Fee, K. Danzmann and S. Chu, \pra 45 (1992) 4911.
\bibitem{Eikema1997} K. S. E. Eikema, W. Ubachs, W. Vassen and W. Hogervorst, \pra 55 (1997) 1866.
\bibitem{Caswell1986} W. E. Caswell and G. P. Lepage, \plb, 167 (1986) 437.
\bibitem{Pachucki2006} K. Pachucki, \pra 74 (2006) 022512.
\bibitem{Stanke2009} M. Stanke, S. Bubin, M. Molski and L. Adamowicz, \pra 79 (2009) 032507.
\bibitem{Bubin2010} S. Bubin, M. Stanke, M. Molski and L. Adamowicz, \cpl 494 (2010) 21.
\bibitem{Rank1962} D. H. Rank, B. S. Rao, P. Sitaram, A. F. Slomba and T. A. Wiggins, \josa 52 (1962) 1004.
\bibitem{Rasetti1929} F. Rasetti, \pr 34 (1929) 367.
\bibitem{Stoicheff1957} B. P. Stoicheff, \cjp 35 (1957) 730.
\bibitem{Crawford1953} M. F. Crawford and I. R. Dagg, \pr 91 (1953) 1569.
\bibitem{May1961} A. D. May, V. Degen, J. C. Stryland and H. L. Welsh, \cjp 39 (1961) 1769.
\bibitem{May1964} A. D. May, G. Varghese, J. C. Stryland and H. L. Welsh, \cjp 42 (1964) 1058.
\bibitem{Looi1978} E. C. Looi, J. C. Stryland and H. L. Welsh, \cjp 56 (1978) 1102.
\bibitem{Veirs1987} D. K. Veirs and G. M. Rosenblatt, \jms 121 (1987) 401.
\bibitem{Germann1988} G. J. Germann and J. J. Valentini, \jpc  92 (1988) 3792.
\bibitem{Rahn1990} L. A. Rahn and G. J. Rosasco, \pra  41 (1990) 3698.
\bibitem{Buijs1971} H. L. Buijs and H. P. Gush, \cjp 49 (1971) 2366.
\bibitem{Durie1960} R. A. Durie and G. Herzberg, \cjp 38 (1960) 806.
\bibitem{Brannon1968} P. J. Brannon, C. H. Church and C. W. Peters, \jms 27 (1968) 44.
\bibitem{McKellar1973} A. R. W. McKellar, \cjp 51 (1973) 389.
\bibitem{McKellar1976} A. R. W. McKellar, W. Goetz, and D. A. Ramsay, \apj 207 (1976) 663.
\bibitem{Rich1982} N. H. Rich, J. W. C. Johns, and A. R. W. McKellar, \jms 95 (1982) 432.
\bibitem{Nazemi1983} S. Nazemi, A. Javan and A. S. Pine, \jcp 78 (1983) 4797.
\bibitem{McKellar1978} A. R. W. McKellar and T. Oka, \cjp 56 (1978) 1315.
\bibitem{Jennings1986} D. E. Jennings, A. Weber and J. W. Brault, \ao 25 (1986) 284.
\bibitem{Bertone2005} G. Bertone, D. Hooper and J. Silk, Phys. Rep. 405 (2005) 279.
\bibitem{Peebles2003} P. J. E. Peebles and B. Ratra, \rmp 75 (2003) 559–606.
\bibitem{Karshenboim2010} S. G. Karshenboim, \prd 82 (2010) 073003.
\bibitem{Ubachs2013} W. Ubachs, W. Vassen, E. J. Salumbides and K. S. E. Eikema, Ann. Phys. (Berlin) 575 (2013) A113. 
\bibitem{Salumbides2013} E. J. Salumbides, J. C. J. Koelemeij, J. Komasa, K. Pachucki, K. S. E. Eikema, and W. Ubachs, \prd 87 (2013) 112008.
\bibitem{Koelemeij2007} J. C. J. Koelemeij, B. Roth, A. Wicht, I. Ernsting, and S. Schiller, \prl 98 (2007) 173002 .
\bibitem{Koelemeij2012} J. C. J. Koelemeij, D. W. E. Noom, D. de Jong, M. A. Haddad and W. Ubachs, Appl. Phys. B 107 (2012) 1075.
\bibitem{Bressel2012} U. Bressel, A. Borodin, J. Shen, M. Hansen, I. Ernsting and S. Schiller, \prl 108  (2012) 183003.
\bibitem{Salumbides2014} E. J. Salumbides, W. Ubachs and V. I. Korobov, submitted to \jms (2014).

\end{thebibliography}
\end{document}